\documentclass[11pt]{JHEP3}
\usepackage{graphicx}
\usepackage{amsfonts}
\usepackage{amssymb}
\usepackage{amsmath}
\usepackage{epsfig}
\newcommand{\be}{\begin{equation}}
\newcommand{\ee}{\end{equation}}
\newcommand{\bea}{\begin{eqnarray}}
\newcommand{\eea}{\end{eqnarray}}
\newcommand{\4}{{\,4}}
\newcommand{\2}{{\,2}}
\title{Anomaly-Mediated Supersymmetry Breaking Demystified}
\author{Dong-Won~Jung\\
Physics Department and CMTP, National Central University,\\
Jhongli, 32054, Taiwan.\\
E-mail: \email{dwjung@ncu.edu.tw}}
\author{Jae~Yong~Lee\\
School of Physics, KIAS,\\
Dongdaemun-Gu, Seoul 130-722, Korea\\
E-mail: \email{littlehigg@kias.re.kr}}
\abstract{We reinterpret anomaly-mediated supersymmetry breaking from a field-theoretic perspective
in which superconformal anomalies couple to either the chiral compensator or the $U(1)_R$ vector superfield.
As supersymmetry in the hidden sector is spontaneously broken by non-vanishing vacuum expectation values
of the chiral compensator F-term and/or the $U(1)_R$ vector superfield D-term,
the soft breakdown of supersymmetry emerges in the visible sector.
This approach is physically more understandable compared with the conventional approach
where the chiral compensator is treated on the same footing as a spurion in gauge-mediated
supersymmetry breaking scenario.}

\preprint{KIAS-P09007}

\begin{document}

\section{Introduction}

The intensive study for the TeV scale physics is just around the corner thanks to the LHC.
Many candidates of new physics beyond the Standard Model are about to be put
to test as the LHC spawns a bonanza of new experimental data in the coming years.
Soon after the first analysis arrives, many candidates will be immediately ruled out
but some will survive the test. This procedure goes on as the opreation of the LHC continues. 
Supersymmetry is the most promising candidate to survive the LHC test.
There are more reasons to support supersymmetry as the LHC's choice.
Supersymmetry $(i)$ provides a promising solution to the hierarchy problem, $(ii)$
explains unification of gauge coupling constant, and $(iii)$ encompasses a candidate
for furtive dark matter of the Universe.
All the three issues are the most urgent problems in particle physics.

Without question, supersymmetry is not found at least up to weak energy scale.
It is believed tht supersymmetry is spontaneously broken in the hidden sector
at some high energy scale and then its breaking is mediated to the visible sector,
leading to the soft terms of the MSSM.
Three prominent mechanisms for the mediation of supersymmetry breaking are
{\it Planck-scale-mediated}~\footnote{It is also referred to as {\it gravity-mediated},
which is an improper use of terminology.}, {\it gauge-mediated} 
and {\it anomaly-mediated} supersymmetry breaking (PMSB, GMSB, and AMSB).
In this article we concentrate on AMSB scenario.
Here the anomaly refers to {\it superconformal} anomaly and the history of its study
goes back to mid 1970's along with developments in rigid supersymmetry theories~\cite{Ferrara:1974pz}.

Though superconformal symmetry is classically preserved in theories without dimensionful parameters
it is in general broken by the quantum effects. Interestingly, the resultant superconformal anomalies
constitute distinct types of supermultiplet by themselves.
There are at least two kinds of superconformal anomaly supermultiplets
- the chiral anomaly and linear anomaly. The former contains three anomalies
- trace anomaly, $U(1)_R$ anomaly and local conformal supersymmetry anomaly
while the latter encompasses only two anomalies (along with a conserved current)
- trace anomaly and local conformal supersymmetry anomaly.
Each anomaly supermultiplet couples to a distinctive superfield in Einstein supergravity.
Though the superconformal anomalies do not break supersymmetry by themselves, 
the coupling of the superconformal anomalies to the superfields in supergravity play
as a mediator of supersymmetry breaking from the hidden sector to the visible sector. 

In the conventional AMSB scenario~\cite{Randall:1998uk,Giudice:1998xp} based on the lagrangian
of the coupled $N=1$ matter-supergravity system in ref.~\cite{Cremmer:1982wb,Cremmer:1982en},
{\it the chiral compensator}\footnote{It is also referred to as {\it the Weyl compensator}
or {\it the conformal compensator}.} plays a central role not only 
as the mediator of supersymmetry breaking but also as the superfield spurion
whose auxiliary component sets the supersymmetry-breaking mass scale.
It is distinguishable from the GMSB scenario where the two roles are played separately
by the two different superfields - the goldstino spurion and gauge messenger(s).
On the other hand, in PMSB scenario the superfield spurion is played by a SM-singlet
superfield of the hidden sector.

In this article we interpret the chiral compensator as a supersymmetric extension
of the Nambu-Goldstone boson (NGB) associated with broken superconformal symmetry.
With this interpretation we understand that the compensator not only plays
the two distinctive roles simultaneously but also is regarded as the invariant measure
of chiral superspace in supergravity.
This leads to the coupling of the chiral compensator to the chiral anomaly
supermultiplet~\cite{Ferrara:1974pz} through the quantum effects.
When the F-term vacuum expectation value (VEV) of the chiral compensator is turned on
by the supersymmetry breaking in the hidden sector~\cite{Pomarol:1999ie},
the soft breaking of supersymmetry in the visible sector emerges 
through the chiral anomaly supermultiplet.
Therefore the soft parameters of the visible sector are characterized both by the F-term VEV
and by the beta function associated with superconformal anomalies.

As the chiral anomaly supermultiplet couples to the chiral compensator 
the linear anomaly supermultiplet~\cite{Sohnius:1981tp,Akulov:1976ck,West} also
couples to a superfield which is referred to as {\it the $U(1)_R$ vector superfield}.
This concurs with a supersymmetrization of the coupling of $R$-current to the $U(1)_R$ gauge field.
In a similar manner to the chiral compensator,
the $U(1)_R$ vector superfield forms a genuine vector superfield.
When the D-term of the $U(1)_R$ vector superfield gets a non-vanishing VEV due to
the hidden sector~\cite{Pomarol:1999ie},
the soft terms in the visible sector take place as well.

The article is organized as follows. In Section 2, we briefly review key ideas
on the connection between conformal symmetry and gravity. 
In Section 3, we set up the chiral anomaly supermultiplet and chiral compensator,
and then find the coupling between them.
In Section 4, we establish the linear anomaly supermultiplet and $U(1)_R$ vector superfield,
and then acquire the coupling between them.
In Section 5, the connection between the anomaly supermultiplet and supercurrent is demonstrated 
with the massless interacting Wess-Zumino model.
In Section 6, we explain how to spawn the soft terms of the visible sector
through the AMSB mechanism.
In Section 7, we summarize the article with the outlook for the future works.
Finally, there are four Appendices which elucidate how to evaluate 
the (super)conformal anomalies using (superspace) perturbation theory and (super)graphs.
 
\section{Conformal symmetry and gravity}%

Let us consider the conformal symmetry in massless QCD.
If all the quarks in QCD were massless then the theory would contain no dimensional parameters
in the lagrangian,
\be {\cal L}=-\frac{1}{4}\,F^a_{mn}F^{amn}+i\sum_{j=1}^F\bar q_jD\!\!\!\! / \,q_j, \ee
and it would exhibit a classical scale invariance.
For conformal (or scale) transformations $x \to e^\varrho x$  where $\varrho$ is arbitrary but real,
the associated quark and gluon scale transformations would be 
\bea q_j(x)&\to& e^{3\varrho/2}q_j(e^\varrho x),\\
A^a_m(x)&\to& e^{\varrho} A^a_m(e^\varrho x).
\eea
These lead to a traceless energy-momentum tensor, with conserved dilatation current 
$J^m_{\rm scale}$,
\be J^m_{\rm scale}=x_n T^{mn},\quad \partial_m J^m_{\rm scale}=T^n_n=0, \ee
where $T^{mn}$ is the energy-momentum tensor.

But conformal symmetry is broken by the quantum effects.
For instance, dimensional regularization imperatively brings about a 
renormalization mass scale $\mu$
which naturally introduces a dimensionful parameter to the theory, 
and therefore breaks conformal symmetry. 
Whenever a global symmetry is spontaneously broken the associated NGB takes place.
The NGB for the broken conformal symmetry is referred to as {\it the dilaton}, denoted by $\varrho$
from now. The dilaton couples to the trace of energy-momentum tensor $T^m_m$,
leading to an effecive action (see Appedix A for how to evaluate it):
\bea\label{eq:qcd.ngb1}
{\check S}_{\rm eff} & = & \int d^\4 x\varrho\, T^m_m,\\
T^m_m &=& \frac{\beta_{\rm QCD}(g)}{2g}\,F^a_{nl}F^{anl},
\eea
where $\beta_{\rm QCD}$ is the beta function for QCD. 
$T^m_m$ is referred to as {\it the conformal} (or {\it trace) anomaly}. 
The effective action is regarded as the coupling of a source (conformal anomaly)
to an external field (dilaton) from a field-theoretic viewpoint.

The effective action (\ref{eq:qcd.ngb1}) can be viewed in a different perspective.
If the gluon field is rescaled to $A^a_m\to gA^a_m$, 
then the effective action is written as
\be\label{eq:qcd.ngb2} \int d^\4 x \varrho\,F^a_{mn}F^{amn}. \ee
That is, renormalization of the gluon field turns the effective action into
the coupling of the dilaton to the lagrangian for the YM theory.
As {\it flat} space is extended to {\it curved} space, the global parameter $\varrho$
in flat space transmutes into the dynamical field $\varrho(x)$ in curved space.
Indeed $\varrho$ in (\ref{eq:qcd.ngb1}) is an integral measure in Einstein gravity
where gravity couples to energy-momentum tensor. 
It is the crossing point where Einstein gravity and quantum field theory without gravity
confront in order to deal with conformal symmetry.

When gravity embraces supersymmetric matter, Einstein gravity is extended to {\it Einstein supergravity}.
Likewise, conformal symmetry is extended to {\it superconformal symmetry}.
Nonetheless conformal symmetry is broken at the quantum level and so is superconformal symmetry. 
As broken conformal symmetry generates a dilaton, broken superconformal symmetry brings about
not only a dilaton but also its superpartner, dilatino. Moreover, both the dilaton and dilatino
constitute a supermultiplet. We have already shown that a dilaton couples to quantum conformal anomaly.
In the following two sections, we will extend the coupling of a dilaton to conformal anomaly to superspace
- quantum superconformal anomalies form a supermultiplet and couples to the dilaton superfield.
We will analyze two kinds of superconformal anomaly supermultiplets
and then identify the corresponding dilaton superfields.

\section{Chiral anomaly supermultiplet and chiral compensator}%

In conformal supergravity~\cite{Kaku:1978nz}
all the relevant dynamical contents are encoded in a single gravitational superfield
({\it i.e.} in the Wess-Zumino gauge) as 
\be\label{eq:grav:field}
{\cal H}^m(x,\theta,\bar\theta)=
\theta\sigma^a\bar\theta e^m_a(x)+\frac{i}{2}\bar\theta\bar\theta\theta \psi^m(x)
-\frac{i}{2}\theta\theta\bar\theta\bar\psi^m(x)
+\frac{1}{4}\theta\theta\bar\theta\bar\theta \hat\nu^m(x).
\ee
Here $\hat\nu^m(x), e^m_a(x)$ and $\psi^m_\alpha(x)$ are a $U(1)_R$ gauge transformations,
vierbein and gravitino, respectively, and their degrees of freedom are three, five and eight, respectively.
As for $\hat\nu^m$ we remove one degree of freedom for {\it the local $R$-symmetry} out of four
so the three degrees of freedom turn out to form the $U(1)_R$ gauge field. 
As for $e^m_a$ we subtract six degrees of freedom corresponding to local Lorentz invariance,
and four degrees of freedom corresponding to general coordinate invariance, and one
degree of freedom corresponding to dilatations. Similarly, as for $\psi^m_\alpha$ we take off
four degrees of freedom for local supersymmetry generators by ${\cal Q}_\alpha$ and $\bar {\cal Q}_{\dot\alpha}$,
and four degrees of freedom for local conformal supersymmetry generators
${\cal S}_\beta$ and $\bar {\cal S}_{\dot\beta}$.

The conformal supergravity multiplet~\cite{Kaku:1978nz} couples to the supercurrent multiplet~\cite{Ferrara:1974pz}
- energy momentum tensor $T_{mn}$, supersymmetry current $S^m_\alpha$
(along with its conjugate, $\bar S^m_{\dot\alpha}$)\footnote{It is also referred to
as spin-$\frac{3}{2}$ supercurrent in ref.~\cite{Wess}.}  and $R$-current $j^R_m$.
When the supercurrent satisfies the conservation equations
\be\label{eq:conserv} \partial^m j^{R}_m = (\gamma_m S^m)_\alpha=T^m_{\,\,m}=0, \ee
its components form a (real scalar) supermultiplet. 
It can be easily read off from the fact that the degrees of freedom for $j^{R}_m,T_{mn}$ and $S^m_\alpha$
are three, five and eight, respectively.
Recall that the supercurrent is not local but global current. 
Thus in rigid supersymmetry theories the superconformal gravity multiplets are treated
as external fields for the supercurrent (we will exhibit it in Section 5).
We apply the current-external-field relation to superconformal anomalies
and superfield in Einstein supergravity as well.  
 
We suppose that all the three symmetries are anomalous. 
That is, none of the three conservation equations is satisfied:
\be\label{eq:anom.set}
\xi_\alpha\equiv\gamma_m S^m_\alpha\neq 0,\quad
\mathring{t}\equiv T^m_m\neq 0,\quad
\mathring{r}\equiv\partial^m j^R_m\neq 0. \ee
The degrees of freedom for the anomalies $\xi_\alpha, \mathring{t}$ and $\mathring{r}$
turn out to be four, one and one, respectively.
To become a supermultiplet we need two new bosonic degrees of freedom, denoted by $a$ and $b$.
They are auxiliary components.
What's more, the four bosonic degrees of freedom are divided into two distinct groups which
contain two degrees of freedom each. It implies that the supermultiplet is {\it chiral}.
The anomaly is referred to as {\it the chiral anomaly supermultiplet} (CASM)~\cite{Clark:1978jx}:
\be\label{eq:anom.sup}
{\cal X}(x,\theta)\equiv {\cal A}(x)+\sqrt{2}\theta\xi(x)+\theta\theta{\cal F}(x),
\quad \bar D {\cal X}=0,\ee
where ${\cal A}=a+ib$ and ${\cal F}=\mathring{t}+i\mathring{r}$.
In contrast to a genuine chiral superfield which has auxiliary
fields as the $\theta\theta$-component, the CASM places them at the zero-component,
a fact which will have an immediate consequence for supersymmetry breaking.
Since we know the mass dimension of conformal anomaly and/or supersymmetry anomaly, 
($[{\cal F}]=4$, $[\xi_\alpha]=7/2$), we find $[{\cal A}]=[{\cal X}]=3$.
Moreover the trace and $U(1)_R$ anomalies go in pairs~\cite{ArkaniHamed:1997mj}
so as to become complex. For instance, in SYM theory the trace anomaly is given as $F_{mn}F^{mn}$
while the $U(1)_R$ anomaly as $F_{mn}\tilde F^{mn}$.
We will comment on it at the end of this section.  

Let us figure out the superfield which couples to the CASM.
As shown in Section 2, the trace anomaly $\mathring{t}$ should couple to the dilaton, 
$\varrho(x)=1/2\ln\mbox{det}[e^m_a]$.
Similarly, the $U(1)_R$ anomaly $\mathring{r}$ should couple to the degree of freedom
for the local $R$-symmetry.
From a field-theoretic point of view, it is the NGB arising from the broken $U(1)_R$ symmetry.
We denote it as $\delta(x)$ from now.
The supersymmetry anomaly $\xi_\alpha$ should couple to dilatino
which couples to the degrees of freedom for the local conformal supersymmetry generators
(approximately $\bar\Psi_\alpha(x) =\sigma_m^{\alpha\dot\alpha}\bar\psi^m_{\dot\alpha}(x)$).
These all simply reinforce the fact that {\it the Nambu-Goldstone bosons or fermions arising
from broken global symmetries in rigid supersymmetric theories
agree with the degrees of freedoms for broken local symmetries in supergravity.}
Nonetheless we need two more bosonic degrees of freedom to balance the bosonic
and fermionic degrees of freedom. They are not associated with any symmetries
so that they are auxiliary fields that couple to ${\cal A}$
in (\ref{eq:anom.sup})~\cite{Ferrara:1978em,Stelle:1978ye}.
Thus these four bosonic and four fermionic degrees of freedom are accommodated
in a chiral superfield of the form~\cite{Ferrara:1983dh}
\be\label{eq:full.comp}
\chi^3(x,\theta) \equiv  e^{2\varrho(x)+2i\delta(x)}
[1+\sqrt{2}\theta \bar\Psi(x)+\theta\theta M^\ast(x)],\ee
where $M$ is complex. It is interesting to note that the role of the Bose components
in the chiral compensator is reveresed by that of the CASM.
This superfield is referred to as {\it the chiral compensator}.
The mass dimension of the chiral compensator and its components are 
$[\varrho]=[\delta]=[\chi^3]=0,[\bar\Psi_\alpha]=1/2$ and $[M]=1$.

In addition to constituting a chiral superfield,
the chiral compensator plays a role in supergravity. As previously mentioned,
it is eaten by the gravitational superfield $(\psi^m_\alpha, e^m_a, \hat\nu_m)$, 
whose degrees of freedom are thereby increased by four, one, one, respectively. 
Moreover, to be a complete multiplet, the gravitational superfield also contains $M$.

In Einstein supergravity,
the chiral compensator is defined as an invariant measure in chiral superspace
\be d^\4 x'd^\2\theta' \chi'^3(x',\theta')=d^\4 x d^\2\,\theta\chi^3(x,\theta). \ee
We can comprehend this property by decomposing the chiral compensator as
\be
d^\4 xd^\2\theta\chi^3(x,\theta)=
\bigg\{d^\4 x\,e^{2\varrho(x)}\bigg\}\bigg\{d^\2\theta\,e^{2i\delta(x)}[1
+\sqrt{2}\theta^\alpha \bar\Psi_\alpha(x)+\theta\theta M^\ast(x)]\bigg\}.
\ee
Note that the term in the first curly bracket is the invariant measure in general relativity 
without supersymmetry while the first term in the second curly bracket is the integral measure 
in $\theta$-space. $\delta$ is the phase associated with the local $R$ rotations.
By rotating $\theta_\alpha\to \theta_\alpha e^{-i\delta}$
the phase $\delta$ can disappear in $\theta$-space.
But the rotation transforms $\bar\Psi_\alpha$ and $M$, respectively, as
\bea \bar\Psi_\alpha & \to & \bar\Psi_\alpha e^{-i\delta},\\
M & \to & Me^{-2i\delta}, \eea
so the chiral compensator is not uniquely defined.
This illustrates that the chiral compensator is covariant under a gauge group
- the local $R$ rotations. This is a supergravity aspect.
Therefore the {\it local} $R$ rotations in supergravity are anything but
the {\it global} $U(1)_R$ rotations in rigid supersymmetry theories so that 
the CASM components transform under the $U(1)_R$ rotations as
\bea {\cal A} & \to & {\cal A},\\
\xi_\alpha & \to & \xi_\alpha e^{-i\delta},\\
{\cal F} & \to & {\cal F} e^{-2i\delta}.
\eea

We are ready to build up the coupling of the chiral compensator to the CASM.
We utilize the properties of CASM and chiral compensator that we have found so far. 
The mass dimension of the CASM and the chiral compensator are three and zero, respectively,
and the chiral compensator plays a role as an integral measure in chiral superspace.
Moreover, we have already described how the components of the CASM couple to
the components of the chiral compensator in advance.
Thus it is not difficult to figure out an action of the form
\be\label{eq:chi.sup}
S_{\cal X}=\int d^\4 x\,d^\2\theta\,\chi^3(x,\theta) {\cal X}(x,\theta)+ h.c.\ee
For instance, we can come up with the same action by extending the effective action
in (\ref{eq:qcd.ngb2}) to superspace.

Let us view the action from the aspect of effective field theory.
A genuine chiral superfield in rigid supersymmetry theories is achieved by multiplying
the chiral compensator by a mass parameter which is assumed be the Planck mass.
Then the chiral superfield $\varphi$ is explicitly written as
\bea \varphi(x,\theta) & = & M_{\rm pl}\,\chi^3(x,\theta)\nonumber\\
& = & M_{\rm pl} e^{2\varrho+2i\delta}
[1+\sqrt{2}\theta^\alpha \bar\Psi_\alpha+\theta\theta M^\ast].
\eea
At much low energy scale ({\it i.e.} weak scale), the dilaton has no VEV ($\langle\varrho\rangle=0$)
while $\langle\delta\rangle$ is unknown. Therefore the $\theta\theta$ component is given by
$M_{\rm pl} e^{2i\langle\delta\rangle} M^\ast$ at low energy scale.    
In this regard we fathom why $M^\ast$ is a complex auxiliary field.
The action is rewritten as
\be S_{\cal X}=
\int d^\4 x\,d^\2\theta \frac{1}{M_{\rm pl}} \varphi(x,\theta) {\cal X}(x,\theta)+h.c.\ee
It is a nonrenormalizable interaction suppressed by the Planck mass 
due to $[{\cal X}]=3$ and $[\varphi]=1$. 
If $\varphi$ has a non-vanishing VEV,
$\langle \varphi\rangle/M_{\rm pl}=1+\theta\theta \langle M^\ast\rangle$ 
(with $\langle\varrho\rangle=\langle\delta\rangle=0$),
then the action leads to soft terms in the visible sector.
In this regard, we would say that anomaly mediation is regarded as Planck-scale-mediation.

Before ending up this section we would like to make a few comments on the action
with regard to the $U(1)_R$ rotations.
Expanding the action (\ref{eq:chi.sup}) in components, we have
\be\label{eq:anom.act} 
S_{\cal X}=\int d^\4 x\, [e^{2\varrho+2i\delta}(M^\ast{\cal A}+\bar\Psi\xi+{\cal F})+h.c.].
\ee
Let us consider the action in the limit $\varrho,\delta\to 0$.
The first term leads to soft supersymmetry breaking terms in the visible sector
if $M$ acquires a non-vanishing VEV. The second term represents
the coupling of the dilatino (gravitino) to the supersymmetry anomaly.
The third term is associated with both conformal and $U(1)_R$ anomalies -
the conformal anomaly survives while
the $U(1)_R$ anomaly fade away.
The more interesting setup is that $\delta$ has a non-vanishing VEV,
$\langle \delta \rangle \neq 0$, regardless of $\varrho$.
It triggers $U(1)_R$ anomalies which are proportional to $\langle \delta\rangle$.
For instance, in YM theory it brings in the $F_{mn}\tilde F^{mn}$ term.
The most interesting situation is that $\delta$ is a dynamical field like $\varrho$.
It is anything but {\it an axion} arising from anomaly mediation.
We will not discuss it in this article.
Studies on phenomenologies of the axion arising from anomaly mediation
under current investigation~\cite{jaeyong}. 
 
\section{Linear anomaly supermultiplet and $U(1)_R$ vector superfield}%

In this section we will address a new superconformal anomaly (superfield).
To begin with, let us return to the three conservation equations (\ref{eq:conserv}).
We now assume that only $R$-current is still conserved but the other two currents become anomalous,
\bea &&\partial^m j^R_m=0\label{eq:r.curr},\\
&&\xi_\alpha=\gamma_m S^m_\alpha\neq 0,\quad \mathring{t}=T^m_m\neq 0, \eea
and then the degrees of freedom for $\mathring{t}$ and $\xi_\alpha$ are one and four, respectively.
To constitute a superfield, three new bosonic degrees of freedom are needed.
$j^R_m$ is the conserved current for the global $U(1)_R$ transformations
in rigid supersymmetric theories (or for the local $U(1)_R$ gauge transfortions in supergravity).
Moreover, the degrees of freedom for $j^R_m$ are three, as shown in Section 3.
Taking the two hints, we can figure out that the new three bosonic degrees of freedom 
are {\it indeed} $j^R_m$. 
This in turn verifies that the anomaly supermultiplet is {\it linear}.
It is referred to as {\it the linear anomaly supermultiplet} (LASM)\footnote{It contains
the same degrees of freedom with {\it the tensor multiplet}~\cite{Siegel:1979ai} where
an antisymmetric-tensor gauge field takes the place of the gauge field in the linear superfield.}
in ref.~\cite{Sohnius:1981tp,Akulov:1976ck}.

Let us identify the LASM components with its corresponding degrees of freedom.
Using the most general expression for a real linear superfield, we can write down
the LASM in terms of its power series expansion in $\theta$
and $\bar\theta$~\cite{Ferrara:1983dh}:
\bea\label{eq:lin.an.fi}
L(x,\theta,\bar\theta)&=&C(x)+i\theta\Xi-i\bar\theta\bar\Xi
+\theta\sigma^m\bar\theta j^R_m(x)\nonumber\\
&&-\frac{1}{2}\,\theta\theta\bar\theta\bar\sigma^m\partial_m\Xi(x)
-\frac{1}{2}\,\bar\theta\bar\theta\theta\sigma^m\partial_m\bar\Xi(x)
-\frac{1}{4}\,\theta\theta\bar\theta\bar\theta\, \Box C(x).
\eea
All the components are expressed by $C$, $\Xi$ and $j^R_m$.
Moreover, the $R$-current is assigned to the $\theta\bar\theta$-component, as expected.
It is interesting to note that LASM has the field content of a vector superfield.
However the roles of the Bose components of the LASM, compared with a vector field,
are reversed: the scalar is now physical (instead of auxiliary),
and the transverse vector now has one physical component and two auxiliary components
(instead of two physical and one auxiliary).
With a knowledge of $[j^R_m]=3$ we read off the mass dimension of its components
- $[L]=[C]=2$, and $[\Xi]=5/2$.
Therefore we can identify the degrees of freedoms for $\mathring{t}$ (conformal anomaly)
and $\xi_\alpha$ (supersymmetry anomaly) with $\Box C$ and
$\sigma^m_{\alpha\dot\alpha}\partial_m\bar\Xi^{\dot\alpha}$, respectively.

Let us look for the superfield corresponding to the LASM.
From a field-theoretic perspective, there must be an external vector field
which couples to the $R$-current. On the other hand, gravitational superfield encompasses
the $U(1)_R$ vector field as components, which is explained in Section 3 
(when we count the degrees of freedom for the gravitational superfield (\ref{eq:grav:field})).
Thus the $U(1)_R$ vector field, $\nu_m$, is transmuted into
{\it the $U(1)_R$ vector superfield}, ${\cal V}$.
Using the following supersymmetric generalization of $U(1)_R$ gauge transformations:
\be\label{eq:u1group} {\cal V}\to{\cal V}+\Lambda+\Lambda^+, \ee
with $\Lambda$ being a chiral field $(\bar D \Lambda=0)$,
we construct the most general expression for ${\cal V}$
written in terms of its power series expansion in $\theta $ and $\bar\theta$:
\bea\label{eq:full}
{\cal V}(x,\theta,\bar\theta)&=&s(x)+i\theta \omega(x)-i\bar\theta\bar\omega(x)\nonumber\\
&&+\frac{i}{2}\,\theta\theta[p(x)+iq(x)]-\frac{i}{2}\,\bar\theta\bar\theta[p(x)-iq(x)]\nonumber\\
&&-\theta\sigma^m\bar\theta \nu_m(x)+i\theta\theta\bar\theta[\bar\tau(x)
+\frac{i}{2}\,\bar\sigma^m\partial_m\omega(x)]\nonumber\\
&&-i\bar\theta\bar\theta\theta[\tau(x)+\frac{i}{2}\,\sigma^m\partial_m\bar\omega(x)]
+\frac{1}{2}\,\theta\theta\bar\theta\bar\theta[{\mathbf d}(x)+\frac{1}{2}\,\Box s(x)].
\eea
Checking out mass dimension of its components, we come up with $[{\cal V}]=[s]=0$,
$[\nu_m]=1$, $[\tau_\alpha]=3/2$ and $[{\mathbf d}]=2$. In the Wess-Zumino gauge
where $s=\omega=p=q=0$, we identify the remaining (physical) components -
$\nu_m$, $\tau_\alpha$ and ${\mathbf d}$ are the gauge field, gaugino, and $D$-term,
respectively, for $U(1)_R$ gauge group.
Fixing this gauge breaks supersymmetry but still allows $U(1)_R$ gauge transformations
\be \label{eq:vec.trans} \nu_m  \to  \nu_m -i\partial_m(\Lambda |-\Lambda^+|).\ee
Note that ${\mathbf d}$ and $\tau_\alpha$ are gauge invariant component fields in ${\cal V}$,
and must be connected to the degrees of freedom for the dilaton and dilatino, respectively.
However, as we learned from Section 3, the mass dimensions of the dilaton and dilatino are,
respectively, zero and 1/2 while $[{\mathbf d}]=2$ and $[\tau_\alpha]=3/2$.
The discrepancy addresses a question - how the components of the $U(1)_R$ vector superfield
are associated with the dilaton and dilatino. To answer the question, we should uncover
the relation between the $U(1)_R$ vector superfield and the chiral compensator. 

Let us delve into the components of ${\cal V}$.
We start with the lowest component or $s$ in eq.~(\ref{eq:full}).
From $U(1)_R$ gauge transformations (\ref{eq:vec.trans}) it is clear that
$(\Lambda |-\Lambda^+|)$ is proportional to the phase of $U(1)_R$ rotations, $\delta$.
Therefore it is not difficult to expect that $(\Lambda |+\Lambda^+|)$ in (\ref{eq:u1group})
is proportional to the dilaton, $\varrho$, in that the chiral compensator contains
the two fields as the scalar component (see eq.~(\ref{eq:full.comp})). 
Thus we find that $s$ is proprotional to $\varrho$.
Eventually as a consequence of the extended gauge transformations (\ref{eq:u1group})
we propose the equivalent of ${\cal V}$~\cite{deWit:1981fh}:
\be {\cal V}\sim \ln [(\chi^3)^+\chi^3], \ee
where $\chi^3$ is the chiral compensator in Section 3.
Why we name the relation as an equivalent (instead of an equality) is evident
if we insert the explicit expression for $\chi^3$ into the log expression.
One cannot achieve the full expression (\ref{eq:full}). This is the downside of
presenting ${\cal V}$ in terms of $\chi^3$. However, this replacement is enough for us
to identify the physical components of ${\cal V}$:
\bea  \nu_{\alpha\dot\alpha} &=& \Psi_\alpha\bar\Psi_{\dot\alpha},\\
\tau_\alpha &=& \bar\Psi_\alpha M,\\
{\mathbf d} &=& M^\ast M, \eea
where $\nu_{\alpha\dot\alpha}=\nu_m \sigma^m_{\alpha\dot\alpha}$.
Evantually with these equalities 
we resolve the discrepancy of the mass dimensions between
the dilaton (dilatino $\Psi_\alpha$) and ${\mathbf d}$ ($\tau_\alpha$).

We will construct the coupling of the $U(1)_R$ vector superfield to the LASM.
There are three conditions that the coupling satisfies: $(i)$ it should be invariant
under the extended gauge transformations, $(ii)$ it should encompass
the coupling of the $R$-gauge field to the $R$-current, and
$(iii)$ the role of the Bose components in the LASM is reveresed
by that of the $U(1)_R$ vector superfield in the Wess-Zumino gauge~(note that
this argument is also applied to the coupling of the CASM to the chiral compensator).
The action for the coupling was found in ref.~\cite{deWit:1981fh,Siegel:1979ai}:
\be\label{eq:super.ac2} S_L=\int d^\4 xd^\2\theta d^\2\bar\theta\, {\cal V}L.\ee
Remind that $[{\cal V}]=0$ and $[L]=2$.

We verify the requirements of the action.
First, let us check the invariance of the action under the extended gauge transformations.
The LASM should be invariant under the extended gauge transformations while
the $U(1)_R$ vector superfield transforms as (\ref{eq:u1group}).
The action transforms as
\bea  \int d^\4 xd^\4\theta\, {\cal V}L &\to &
\int d^\4 xd^\4\theta\,({\cal V}+ \Lambda + \Lambda^+)L,\nonumber\\
&=&\int d^\4 xd^\4\theta\,{\cal V}L
+\int d^\4 xd^\2\theta \Lambda(\bar D^2L)+\int d^\4 xd^\2\bar\theta \Lambda^+(D^2 L). \eea
For the action to be invariant under the extended gauge transformations,
the LASM satisfies the two conditions:
\be D^2 L= \bar D^2 L= 0,\ee
which in turn imply that $L$ is indeed {\it the current superfield}.
As we previously mentioned, $L$ is anything but {\it the $R$-current superfield}. 

Let us find out the coupling of the $R$-gauge field to the $R$-current.
As we expand the action in components, we get
\be \label{eq:lin.coup}
S_L=\int d^\4 x\,(\nu^m j^R_m +\frac{1}{2}\,{\mathbf d}\,C+\tau\,\Xi+\bar\tau\,\bar\Xi). \ee
As intended, the first term is the coupling of the $R$-gauge field to the $R$-current, and
is invariant under the gauge transformations (\ref{eq:vec.trans})
due to the $R$-current conservation equation (\ref{eq:r.curr}). 
The second term is the only coupling associated with $C$. 
It seems that the action embraces two more terms involved in $C$,
\be S_L\supset \frac{1}{4}\int d^\4 x\,C(\Box s)-\frac{1}{4}\int d^\4 x\,(\Box C)s. \ee
But they cancel each other by virtue of integration by parts.
This is clear in the Wess-Zumino gauge where $s=0$.
Therefore, $\int d^\4 x\,{\mathbf d}\,C$ is the only term
that exhibits the footprint of conformal anomaly. 
Once ${\mathbf d}$ acquires a non-vanishing VEV which originates
from the hidden sector, the action leads to the soft terms in the visible sector.
The last two terms in (\ref{eq:lin.coup}) reveal the footprint of supersymmetry anomaly. 

\section{Supercurrent, supertrace and anomaly supermultiplet}%

In this section we describe the relation between the anomaly supermultiplets and supertrace.
First of all, we review the relation between supercurrent and 
supertrace in the context of Einstein supergravity.  
Let us consider an action of matter chiral superfields $\phi(x,\theta)$
living on a given (background) curved space with the gravitational superfeld
${\cal H}^{\alpha\dot\alpha}(x,\theta,\bar\theta)\equiv{\cal H}^m(x,\theta,\bar\theta)
\sigma_m^{\alpha\dot\alpha}$
and chiral compensator $\chi(x,\theta)$:
\be S=S[\phi;{\cal H}^{\alpha\dot\alpha},\chi], \ee
and satisfying their dynamical equations
\be \frac{\delta S[\phi;{\cal H}^{\alpha\dot\alpha},\chi]}{\delta\phi}=0. \ee  
The conservation equation leads to the relation
\be \bar {\cal D}^{\dot{\alpha}}J_{\alpha\dot{\alpha}}={\cal D}_\alpha {\cal X}, \ee
where $J_{\alpha\dot{\alpha}}$ and ${\cal X}$ are
supercurrent and supertrace, respectively, defined as
\bea
J_{\alpha\dot{\alpha}}&\equiv&\frac{\delta S}{\delta {\cal H}^{\alpha\dot{\alpha}}},\quad
J_{\alpha\dot\alpha}=J_m\sigma^m_{\alpha\dot\alpha},\quad J_m=J^+_m\\
{\cal X}&\equiv&\frac{\delta S}{\delta \chi^3}, \quad \bar{\cal D}^{\dot\alpha} {\cal X} =0,
\eea
and ${\cal D}_\alpha(\bar{\cal D}_{\dot\alpha})$ is a covariant derivative 
(a local version of $D_\alpha$ in rigid supersymmetry).
This indicates that the supercurrent and supertrace couple to the gravitational superfield
and chiral compensator, respectively.
The supertrace is anything but the CASM.
It is worthwhile to remark that the $supercurrent$ is different
from the $R$-current superfield in Section.4.

If there are no dimensionful parameters in the action then the action is
invariant under the superconformal transformations.
Then the supertrace vanishes on-shell. Namely, ${\cal X}=0$.
Due to the relation the classical supercurrent satisfies the conservation relation
\be \bar {\cal D}^{\dot{\alpha}}J_{\alpha\dot{\alpha}}=0, \ee 
which is equivalent to the superconformal conservation equations~(\ref{eq:conserv}).
Though a theory is classically invariant under superconformal symmetry
the quantum effects break the symmetry.
Even in {\it flat} space, superconformal anomaly takes place at the quantum level.

Let us evaluate quantum superconformal anomaly in flat space ($\chi=1$).
As a toy model, we compute the supercurrent in the massless interacting Wess-Zumino model
described by the action:
\be S=\int d^\4 xd^\4 \theta\,\phi^+\phi
+\bigg[\int d^\4 xd^\2\theta\,\frac{y}{3!}\,\phi^3+h.c.\bigg], \ee
where $\phi(x,\theta)$ is a matter (chiral) superfield and $y$ is a dimensionless Yukawa coupling. 
As mentioned in Section 3, the classical supercurrent contains the three components 
- energy momentum tensor $T_{mn}$, supersymmetry current $S_m$ (along with its conjugate, $\bar S_m$)
and $R$-current $j^R_m$ such as
\be
J_m(x,\theta,\bar\theta)=
j^{R}_m(x)+\theta S_m(x)+\bar\theta \bar S_m(x)+\theta\sigma^n\bar\theta\,T_{mn}(x).
\ee
Using the Noether procedure, the supercurrent is easily constructed as
\be
J_{\alpha\dot\alpha}=-\frac{1}{3}D_\alpha\phi\bar D_{\dot\alpha}\phi^+
+\frac{1}{3}\phi^+ i\overleftrightarrow{\partial}_{\!\!\alpha\dot\alpha}\phi.
\ee
The supercurrent satisfies the conservation equation
\be 
\bar D^{\dot{\alpha}}J_{\alpha\dot{\alpha}}=0,
\ee
by virtue of the equation of motion.
It implies that there is no supertrace and superconformal symmetry is preserved {\it classically}. 
This is because the massless interacting WZ model contains no dimensionful parameters.

Let us jump from classical theory to quantum theory. The supercurrent is replaced
by a renormalized operator defined by computing 1PI Green functions
with one $J_{\alpha\dot\alpha}$ insertion and subtracting divergences.
Quantum effects lead to superconformal anomalies which are easily computed
from the dependence of the effective action on the superconformal compensator of supergravity
(see the ref.~\cite{Grisaru:1983hc,Grisaru:1985ik} for the computation).
The superconformal anomaly at one loop level is described by the relation
\be
\bar D^{\dot{\alpha}}J_{\alpha\dot{\alpha}}=
-\frac{1}{9}\frac{\beta(y)}{y}D_\alpha K,
\ee
where $K$ is the chiral superfield $K=\bar D^2(\phi^+\phi)$,
and $\beta(y)$ is the beta funcion of the coupling $y$,
\be \frac{\beta(y)}{y}=\frac{3}{2}\bigg(\frac{y}{4\pi}\bigg)^2+\cdots.\ee
From the relation we can easily read off the supertrace,
\be {\cal X}=-\frac{1}{9}\frac{\beta(y)}{y} \bar D^2(\phi^+\phi). \ee
Using the equation of motion
\be \bar D^2 \phi^+=-\frac{1}{2}\,y \phi^2,\ee
we can rewrite the CASM as
\be {\cal X}=\frac{\beta(y)}{3y} \frac{y}{3!}\,\phi^3. \ee 
This is nothing but the CASM of the massless interacting WZ model.
Note that it is given as a product of the beta function and the cubic interaction term. 

So far we have taken into account of the CASM. We now turn our attention to the LASM.
We will redefine the supercurrent in such a way that the $R$-current is conserved:
\bea
\tilde J_{\alpha\dot\alpha}&=&J_{\alpha\dot\alpha}-\bigg(\frac{y}{4\pi}\bigg)^2 K_{\alpha\dot\alpha},\\
K_{\alpha\dot\alpha}&=&-\frac{1}{6}[\bar D_{\dot\alpha},D_\alpha](\phi^+\phi).
\eea
The redefined  supercurrent satisfies the following relation at ${\cal O}(y^2)$, 
\be \bar D^{\dot\alpha} \tilde J_{\alpha\dot\alpha}=\frac{1}{2}\bigg(\frac{y}{4\pi}\bigg)^2 \bar D^2 D_\alpha (\phi^+\phi).\ee
This is nothing but the conservation equation involved with the LASM in the curved space:
\be \bar {\cal D}^{\dot{\alpha}}\tilde J_{\alpha\dot{\alpha}}=\bar {\cal D}^2 {\cal D}_\alpha L, \ee
where the LASM is given as
\be L=\frac{1}{2}\bigg(\frac{y}{4\pi}\bigg)^2 (\phi^+\phi). \ee
We note that the LASM is given as a product of the kinetic term and the gamma function at ${\cal O}(y^2)$
\be \gamma= \frac{1}{2}\bigg(\frac{y}{4\pi}\bigg)^2+\cdots.\ee
In the context of Einstein supergravity, it indicates that the background curved space switches
from ${\cal H}^{\alpha\dot\alpha}$ and $\chi^3$ to
$\tilde{\cal H}^{\alpha\dot\alpha}$ (a new gravitational superfield) and ${\cal V}$.

Before closing this section we would like to comment a thing.
In general any finite subtraction on the $R$-current cannot be extended to the whole supercurrent
without destroying the conservation or symmetry properties of the renormoralized energy-momentum tensor.
The massless interacting WZ model is an exception to this rule \cite{Clark:1978jx}.

\section{Anomaly mediated supersymmetry breaking}%

So far we have concentrated on the couplings of superconformal anomalies to their corresponding
superfields in supergravity. With these couplings we are equipped for probing the mediation
of supersymmetry breakdown to the visible sector ({\it i.e.} MSSM).
We assume that supersymmetry is broken in the hidden sector
and then its effects cause both the F-term of the compensator
and the D-term of the $U(1)_R$ vector superfield to acquire non-vanishing VEVs.
Moreover, the VEVs turn out to equal the gravitino mass, $m_{3/2}$.
We do not explain in the article why the VEV is set to be the gravitino mass.
In the followings, the F-term of the chiral compensator is given by $m_{3/2}$
while the D-term of the $U(1)_R$ vector superfield is given as ${\mathbf d}=m^2_{3/2}$.

We take into account a supersymmetric nonabelian gauge theory
as a simplified version of the MSSM.
Let the simple gauge group $G_A$ with a universal gauge coupling constant $g$.
For simplicity, the chiral matter superfields $\phi_i$ belong to the fundamental
representation of $G_A$.
Let us also specify the superpotential to be that of the massless interacting Wess-Zumino model.
The classical action is then expressed as
\be\label{eq:sim.mssm}
S=\int d^\4 \!x\, d^\2\theta d^\2\bar\theta\,{\phi_i}^{\!+} e^{2gV} \phi_i
+\Bigg[\int d^\4 \!x\,d^\2\theta\,\bigg(\frac{1}{4}\,W^{a\alpha} W^a_\alpha
+\frac{1}{3!}\,y^{ijk}\,\phi_i\phi_j\phi_k\bigg)+h.c.\Bigg].
\ee
Since there are no dimensionful parameters in the theory, superconformal symmetry
is classically preserved. But quantum effects break the superconformal symmetry,
as explained in Section 2.

In the previous sections we have already established the formalism for the the coupling of
superconformal anomalies to the corresponding superfields in supergravity.
All we need is to acquire the superconformal anomaly (supermultiplet) of the model.
As anticipated, the superconformal anomalies are given as the same form of the lagrangian.
The CASM associated with the gauge kinetic term is evaluated at one loop (see Appendix B):  
\bea
{\cal X} &=& \frac{\beta(g)}{2g}\,W^{a\alpha} W^a_\alpha\label{eq:CASM1},\\
\beta(g)&=&-\frac{g^3}{16\pi^2}[3C_A(\mbox{adj.})-\sum_iT_A(\phi_i)],
\eea
where $C_A(\mbox{adj.})$ is the quadratic Casimir of the adjoint representation,
and $T_A(\phi_i)$ is the Dynkin index of the representation to which $\phi_i$ belongs.

Plugging the CASM (\ref{eq:CASM1}) into the action (\ref{eq:anom.act}), and opting for the term
$W^{a\alpha} W^a_\alpha |=-\lambda^a\lambda^a$, we easily read off the gaugino mass,
\be
M_\lambda = \frac{\beta(g)}{g} m_{3/2}.
\ee

Next, let us consider the CASM involved in the Yukawa term. 
The CASM is evaluated at one loop (see Appendix C) as
\bea
{\cal X} &=& -\frac{1}{3!}\,(\gamma_i+\gamma_j+\gamma_k)\,y^{ijk}\phi_i\phi_j\phi_k,
\label{eq:CASM2}\\
\gamma_i^j&=& -\frac{1}{32\pi^2}[y^\ast_{ikl} y^{jkl}-4g^2\delta_i^jC_A(\phi_i)],
\eea
where $\gamma_i^i(\equiv\gamma_i)$ is the anomalous dimension of the chiral superfield $\phi_i$,
and $C_A(\phi_i)$ is the quadratic Casimir of the representation to which $\phi_i$ belongs.
Inserting the CASM (\ref{eq:CASM2}) into the action (\ref{eq:anom.act}) we directly get the $A$-term,
\be A_{ijk} =-(\gamma_i+\gamma_j+\gamma_k) y^{ijk}m_{3/2}.\ee

Finally, we take into account the LASM engendered by the kinetic term of the matter superfield.
The LASM is evaluated at two loop (see Appendix D):
\bea L &=& -\frac{1}{2}\,\dot\gamma_i^i\,{\phi_i}^{\!+}\phi_i,\\
\dot\gamma_i^j&=&\frac{\partial\gamma_i^j}{\partial\ln\mu}.
\eea

We insert the LASM into the effective action in (\ref{eq:super.ac2}). Thereafter we identify
$C=-\frac{1}{2}\dot\gamma_i^i\,{\phi_i}^{\!+}\phi_i|$ and replace ${\mathbf d}$ by $m^2_{3/2}$.
The mass of the scalar field $\phi_i|$ is immediately read:
\be m^2_i =\frac{1}{4}\dot\gamma_i^i\,m^2_{3/2}. \ee 

\section{Summay and outlook}%

We reviewed anomaly-mediated supersymmetry breaking scenario in the 
context of rigid supersymmetric field theory,
almost without knowledge of supergravity aspects.
It is revealed that anomaly-mediated supersymmetry breaking is anything but the chiral compensator-,
and $U(1)_R$ vector superfield-mediation in the sense that a gauge messenger mediates
supersymmetry breaking in gauge-mediation theories.
This interpretation is viable because both the chiral compensator and the $U(1)_R$ vector superfield
are identified as supersymmetric extensions of the dilaton arising from broken conformal symmetry,
and couple to the chiral and linear anomaly supermultiplets of the visible sector, respectively.
Spontaneous breakdown of supersymmetry in the hidden sector is communicated to the
visible sector through non-zero vacuum expectation values of the chiral compensator F-term and
the $U(1)_R$ vector superfield D-term.
In this regard, we understood why the auxiliary fields in Einstein 
supergravity play a key role in anomaly mediation.
As a consequence, it is shown that the gaugino masses and trilinear scalar couplings
arise from the chiral anomaly while the sfermion masses do from the linear anomaly.

We illustrated that both the chiral compensator and the $U(1)_R$ vector superfield are contained
in the gravitational superfield of Einstein supergravity.
These fields fluctuate at the Planck scale so that anomaly mediation scenario can be regarded
as a specific example of Planck-scale-mediated theories.
On the other hand, at low energy scale their effects are left only with their non-vanishing
vacuum expectation values, which explains why the chiral compensator
is treated as a spurion superfield in the conventional anomaly mediation scenario.

We also learned that the $U(1)_R$ vector superfield does not come into view
because it is replaced approximately by the chiral compensator.
Actually the chiral compensator and $U(1)_R$ vector superfield are dynamical fields
as like graviton and gravitino in the context of rigid supersymmetry theories.
In this regard, the mechanism for supersymmetry breaking in the context of superconformal symmetry
can be treated by the langauge of rigid supersymmetry theories.
This viewpoint helps understand not only supersymmetry breaking but also supergravity itself.
For instance, the F-term of the chiral compensator and the D-term of the $U(1)_R$ vector superfield
are intertwined with the gravitino mass through super-Higgs mechansim.
Thus we may thoroughly investigate anomaly mediation not only at low energy scale
but also at high energy scale (as high as GUT scale) including the hidden sector.
Taking one step further toward this perspective, 
one can adopt this method to study the connection between supergravity 
and supersymmetry breaking in the hidden sector, and evaluate the soft terms arising 
from gauge- and anomaly-mediation at once.

\section{Acknowledgements}
We are very grateful to Hyun Seok Yang and Sungjay Lee for numerous discussions and helpful comments.
We also thank Kwang Sik Jeong for pointing out the correct expression for the sfermion mass.
DWJ is supported by the Taiwan NSC grants NSC 97-2811-M-008-024 and NSC 96-2112-M-008-007-MY3.

\appendix%

\section{The coupling of dilaton to conformal anomaly in massless QCD}%

In this appendix we evaluate the conformal anomaly of massless QCD. In particular,
our concern is the coupling of the dilaton, $\varrho$, to the conformal anomaly, ${\cal X}$.
Let us rescale the gluon field by $A^a_m\to\frac{1}{g}A^a_m$ so that 
the classical action for the gluon field is given in $x$-space of Euclidean signature by
\bea
\label{eq:app1} S &=&\int d^\4 x\,\frac{1}{4g^2}\,F^a_{mn}F^a_{mn}\nonumber\\
&=& \int \frac{d^\4 p}{(2\pi)^4}\,\frac{1}{2g^2}\,(p^2\delta_{mn}-p_mp_n)A^a_m(-p)A^a_n(p)
+\cdots,
\eea
where the integral variables are switched from $x$-space to momentum-space
in order to carry out loop calculations.
This is because the quantum effects give rise to the conformal anomaly of the theory.
The gluon self-interactions are omitted because our computation is involved only with
the gluon kinetic term.

As one evaluates the quantum effects with loop diagrams, one in general renormalizes
not only the wavefunctions but also the parameters of the lagrangian.
However, the evaluation of the conformal anomaly is remarkably straightforward
if one adopts {\it the background field method} which deals only with
renormalization of the gauge coupling (see the section 16.6 in ref.~\cite{Peskin:book}).
All the Feynman diagrams contributing to the running gauge coupling in the background
field method are shown in fig.~\ref{fig:A1}.
\FIGURE{\epsfig{file=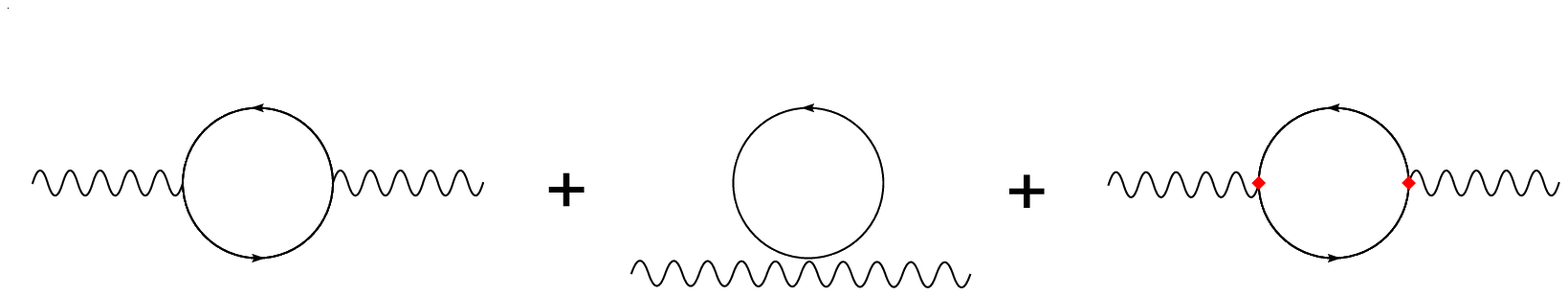,width=0.9\textwidth}
\caption{Terms quadratic in the external gluon field in the expansion of the effective action.
The special vertex arises from the most right diagram. See the page 539 in ref.~\cite{Peskin:book}
for details.} \label{fig:A1}}%
 
Using dimensional regularization in $n\,(=4-2\varepsilon)$ dimensions,
we find the one-loop effective action:
\be\label{eq:eff.act}
S_{\rm eff}=\int \frac{d^\4 p}{(2\pi)^4}\,\frac{1}{2}\,
(p^2\delta_{mn}-p_mp_n){\mathbf A}^a_m(-p){\mathbf A}^a_n(p)
\frac{b}{16\pi^2}\Sigma(p^2),
\ee
where ${\mathbf A}^a_n$ is the gluon background (or external) field, and
the first coefficient $b$ of the beta function is
\bea
\beta(g) &=& -\frac{bg^3}{16\pi^2}+{\cal O}(g^5),\\
b &=& \frac{11}{3}C_A(\mbox{adj.})-\frac{4}{3}\sum_iT_A(q_i),
\eea
with $C_A(\mbox{adj.})$ being the quadratic Casimir constant of the adjoint representation,
$T_A(q_i)$ being the Dynkin index of the representation of the gauge group to which
the quark field $q_i$ belongs.\footnote{The subscript $A$ represents
the distinctive gauge group for multi-gauge theories.}  
Note that $C_A(\mbox{adj.})=N$ and $T_A(q_i)=1/2$ for the $SU(N)$ gauge group
and the quark with the fundamental representation.

It is worthwhile to remark that the beta function is the same with that obtained
by the relation (in the conventional dimensional renormalization method)
\bea \beta(g) &=& \frac{\partial g}{\partial \ln \mu },\\
g_0 &=& g\mu^\varepsilon \frac{Z_1}{Z_2Z_3^{1/2}}, \eea
where $g_0$ is bare coupling, $Z_1$ is the quark-gluon-gluon vertex renormalization
constant, $Z_2$ is the quark wavefuntion renormalization constant, and $Z_3$ is the
gluon wavefuction renormalization constant.

The expression for $\Sigma(p^2)$ is given as
\bea
\Sigma(p^2) &=& (4\pi)^2\mu^{-2\varepsilon}\int \frac{d^{\,n}q}{(2\pi)^n}\,\frac{1}{q^2(p+q)^2}\nonumber\\
&=&\bigg(\frac{p^2}{4\pi\mu^2}\bigg)^{\!-\varepsilon}\frac{\Gamma(\varepsilon)[\Gamma(1-\varepsilon)]^2}
{\Gamma(2-2\varepsilon)}\nonumber\\
&=& \frac{1}{\varepsilon}\bigg[1+\varepsilon\bigg(2-\gamma_E-\ln\frac{p^2}{4\pi\mu^2}\bigg)
+\,{\cal O}(\varepsilon^2)\bigg].
\eea
Note that the factor $\mu^{2\varepsilon}$ comes from the relation between bare
and renormalized couplings, and $\Gamma$ is given as
\be \Gamma(1+\varepsilon)=1-\varepsilon\gamma_E+\varepsilon^2\hat\delta+{\cal O}(\varepsilon^2),\ee
where $\gamma_E$ is the Euler constant, and $\hat\delta=\frac{1}{12}\pi^2+\frac{1}{2}\gamma^2_E$.
Thus the effective action is written by
\be\label{eq:eff.act2}
S_{\rm eff}=\int \frac{d^\4 p}{(2\pi)^4}\,\frac{1}{2g^2}\,
(p^2\delta_{mn}-p_mp_n){\mathbf A}^a_m(-p){\mathbf A}^a_n(p)
\frac{bg^2}{16\pi^2} \bigg(\frac{\mu^2}{p^2}\bigg)^{\!\!\varepsilon}
\frac{1}{\varepsilon}.
\ee

Scale transformations in loops are anything but the shifts in {\it renormalization mass scale},
$\mu \to \mu'$. Since scale transformations in $x$-space is set to be $x\to e^\varrho x$
for arbitrary $\varrho$, transformations of renormalization mass scale is then given by 
\be \mu\to e^{-\varrho}\mu. \ee
Recall that $\varrho$ is a dimensionless parameter at this moment.

We now replace $\mu$ in eq.~(\ref{eq:eff.act2}) by $e^{-\varrho}\mu$
and then single out the linear term in $\varrho$. Regarding $\varrho$ as a dynamical field (dilaton)
from now, we immediately find the coupling of the dilaton to the (quantum) conformal anomaly:
\bea
\check{S}_{\rm eff} &=&
\frac{g^2b}{16\pi^2}\,\frac{1}{2}\int\frac{d^\4 p}{(2\pi)^\4 }\,
\frac{(-2\varrho)}{g^2}(p^2\delta_{mn}-p_mp_n){\mathbf A}^a_m(-p){\mathbf A}^a_n(p)\nonumber\\
&= &-\frac{g^2b}{32\pi^2}\int d^\4 x\,\varrho \,\frac{1}{g^2}\,{\mathbf F}^a_{mn}{\mathbf F}^a_{mn}.
\eea
Thus the coupling is written by rescaling $A^a_m\to gA^a_m$:
\be\check{S}_{\rm eff}=\frac{\beta_{\rm QCD}}{2g} \int d^\4 x\,\varrho \,F^a_{mn}F^{amn},\ee
from which the conformal anomaly is direclty gleaned: 
\be\label{eq:qcd.chiral.ano} {\cal X}=\frac{\beta_{\rm QCD}}{2g}F^a_{mn}F^{amn}. \ee  

One can derive the conformal anomaly using
either a path integral treatment by Fujikawa~\cite{Fujikawa:1979ay} or the calculation of
a Feynman diagram, {\it i.e.} the triangle diagram.
In comparison with other methods for evaluating (quantum) conformal anomalies,
the advantage of employing this method is to make no use of either conformal
(or dilatation) current or energy-momentum tensor.
This method will be adopted for the evaluation of {\it superconformal} anomalies in the following
Appendices. 

\section{The chiral anomaly from the gauge kinetic term}%

In this appedix we evaluate the CASM coming from the gauge kinetic term
of the action (\ref{eq:sim.mssm}) in Euclidean $x$-space~\cite{Grisaru:1979wc}.
We can directly evaluate the anomaly
using superspace perturbation theory and supergraphs. Instead, we indirectly and effortlessly
evaluate the anomaly using the analogy between this theory and massless QCD in Appendix A.

The CASM stemming from the gauge kinetic term,
$1/4(W^{a\alpha} W^a_\alpha+\bar W^a_{\dot\alpha}\bar W^{a\dot\alpha})$, can be evaluated
in a straightforward way using the relations,
\bea
W^{a\alpha}W^a_\alpha|_{\theta\theta} &=&
\frac{1}{2}F^a_{mn}F^{amn}-2i\lambda^a\sigma^m\partial_m\bar\lambda^a+(D^a)^2
+\frac{1}{2}F^a_{mn}\tilde F^{amn},\\
\bar W^a_{\dot\alpha}\bar W^{a\dot\alpha}|_{\bar\theta\bar\theta} &=&
\frac{1}{2}F^a_{mn}F^{amn}+2i\partial_m\lambda^a\sigma^m\bar\lambda^a+(D^a)^2
-\frac{1}{2}F^a_{mn}\tilde F^{amn}.
\eea
Comparing them with eq.~(\ref{eq:qcd.chiral.ano}) we easily read off the (anti)chiral
anomaly superfield:
\bea
{\cal X} &=& \frac{\beta(g)}{2g}W^{a\alpha} W^a_\alpha,\\
{\cal X}^+ &=& \frac{\beta(g)}{2g}\bar W^a_{\dot\alpha} \bar W^{a\dot\alpha},
\eea
where the beta function of the theory is given at one loop by
\bea 
\beta(g) &=& -\frac{bg^3}{16\pi^2}+{\cal O}(g^5),\\
b &=& 3C_A(\mbox{adj.})-\sum_i T_A(\phi_i).
\eea

The beta function can be directly evaluated by the wavefuntion renormalization of the 
vector superfield with the knowledge of
\be Z_gZ_V^{1/2}=1, \ee
where $Z_g(Z_V)$ is the renormalization constant for the gauge coupling (the vector
superfield). See fig.~\ref{fig:B1} for the relevant one-loop supergraphs.
\FIGURE{\epsfig{file=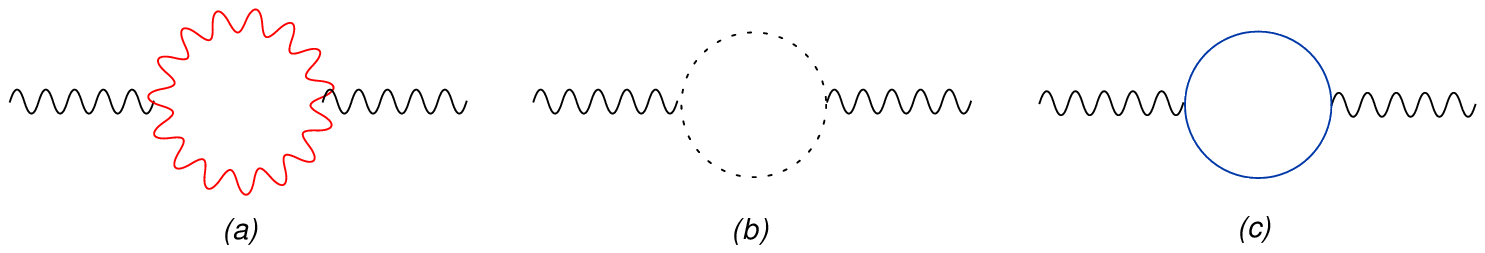,width=0.9\textwidth}
\caption{Supergraphs:~(a)~vector,~(b)~ghost,~and~(c)~chiral one-loop
contribution to the vector superpropagator.}\label{fig:B1}}

\section{The chiral anomaly from the Yukawa interaction}%

In this appendix we evaluate the CASM arising from the Yukawa interaction
of the action (\ref{eq:sim.mssm}) in Euclidean $x$-space~\cite{Grisaru:1979wc}:
\be S \supset - \int d^\4 x\,d^\2\theta\,\frac{1}{3!}\,y^{ijk}\,\phi_i\phi_j\phi_k + h.c.\ee
Recall that the sign is flipped due to Euclidean $x$-space.
Using dimensional regularization we evaluate the one-loop effective action from the diagrams
in fig.~\ref{fig:C1}:
\be\label{eq:eff.act3}
S_{\rm eff}=-\int d^\4 x\,d^\2\theta\,\frac{1}{3!}\,y^{ikl}\bigg\{\frac{1}{32\pi^2}
[y^\ast_{irs}y^{jrs}-4g^2C_A(\phi_i)\delta^j_i]\frac{\mu^{2\varepsilon}}{2\varepsilon}\phi_j\phi_k\phi_l
+(cyclic)\bigg\}.
\ee
It is noteworthy that the one-loop effects stem only from the wavefunction renormalization.
One may take into account other one-loop diagrams shown in fig.~\ref{fig:C2}.
But they either vanish (fig.~4(a)) or make no contribution to the superconformal anomaly
(fig.~4(b)). This is the consequence of the ``non-renormalization'' 
theorems for theories of left-chiral superfields.
\FIGURE{\epsfig{file=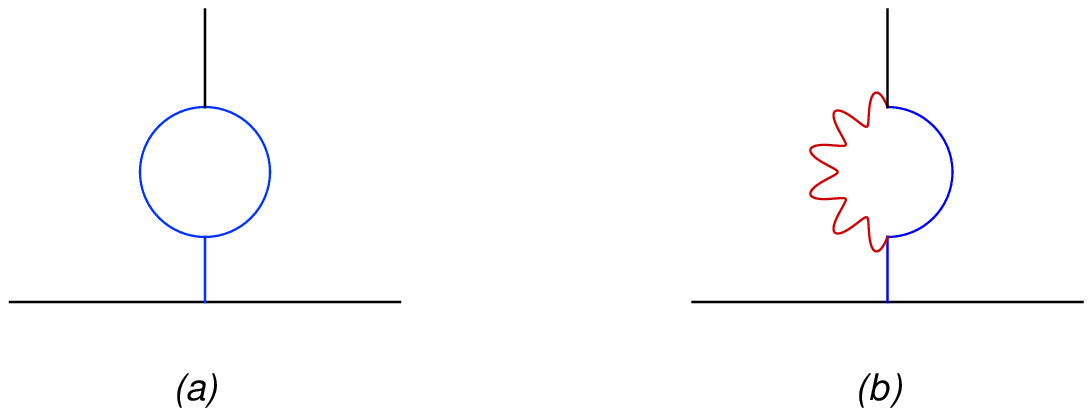,width=0.6\textwidth}
\caption{Supergraphs: one-loop contribution to the superconformal anomaly.
They are log-divergent so that they lead to the wavefunction renormalization.}\label{fig:C1}}

The wavefunction renormalization constant for the matter fields is given as 
(see in Appendix D):
\be Z^{1/2}_i(\mu)=1-\frac{1}{32\pi^2}
[y^\ast_{irs}y^{jrs}-4g^2C_A(\phi_i)]\frac{\mu^{2\varepsilon}}{2\varepsilon},\ee
and the anomalous dimension of the matter superfield $\phi_i$ is given by
\be \gamma_i=-\frac{1}{32\pi^2}[y^\ast_{ijk} y^{ijk}-4g^2C_A(\phi_i)].\ee
Here we assume that all anomalous dimensions are diagonal in field space, {\it i.e.}
$\gamma_{ij}=\gamma_i\delta_{ij}$.

Replacing $\mu$ in eq.~(\ref{eq:eff.act3}) by $e^{-\varrho}\mu$ and then picking up the linear term in $\varrho$,
we get the coupling of the dilaton to the chiral anomaly,
\bea \label{eq:seff3}
\check{S}_{\rm eff}&=&-\int d^\4 x\,d^\2\theta\,\varrho\,\frac{1}{3!}\,y^{ijk}\bigg\{\frac{1}{32\pi^2}
[y^\ast_{irs}y^{irs}-4g^2C_A(\phi_i)]\,\phi_i\phi_j\phi_k+(cyclic)\bigg\}\nonumber\\
&=&\int d^\4 x\,d^\2\theta\,\varrho\,\frac{1}{3!}\,(\gamma_i+\gamma_j+\gamma_k)y^{ijk}\phi_i\phi_j\phi_k.
\eea
The CASM for the Yukawa interaction is then gleaned from the effective action
$\check{S}_{\rm eff}$:
\be
{\cal X} = \frac{1}{3!}\,(\gamma_i+\gamma_j+\gamma_k)\,y^{ijk}\phi_i\phi_j\phi_k.
\ee
Recall that in Minkowski $x$-space the sign of the CASM is flipped.
%%%
\FIGURE{\epsfig{file=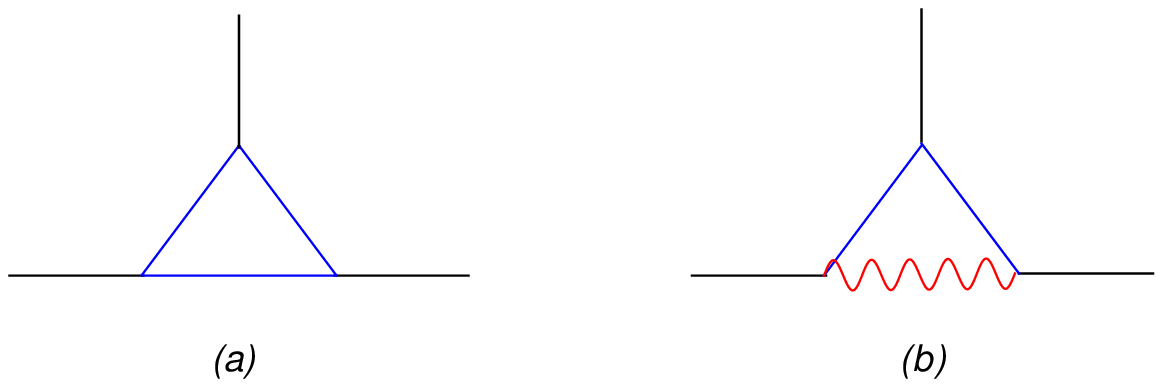,width=0.6\textwidth}
\caption{Supergraphs: one-loop Feynman diagrams that do not contribute to the superconformal anomaly:
the diagram (a) vanishes itself while the diagram (b) is convergent
so that it does not contribute to the anomaly.}\label{fig:C2}}

Taking one step further with this method we supersymmetrize the dilaton. That is, 
the dilaton is transformed into the the chiral compensator:
\be e^{2\varrho} \to \chi^3=
e^{2\varrho+2i\delta}[1+\sqrt{2}\theta\bar\Psi+\theta\theta M^\ast].\ee
Under this transformation, 
\bea \ln\mu^2 &\to & \ln\mu^2 -\ln\chi^3\nonumber\\
&=& \ln \mu^2-2\varrho-2i\delta -\sqrt{2}\theta\bar\Psi-\theta\theta M^\ast.\eea
The combination of $\varrho$ and the trilinear operator $\phi_i\phi_j\phi_k$
in (\ref{eq:seff3}) gives rise to the coupling of the dilaton to the chiral anomaly superfield
while the combination of $\ln \mu^2$ and the trilinear operator $\phi_i\phi_j\phi_k$ yields
the anomalous dimension of the operator $\phi_i\phi_j\phi_k$.
It is noticeable that the combination of $\delta$ and the trilinear operator $\phi_i\phi_j\phi_k$
yields the coupling of the axion to the the trilinear operator $\phi_i\phi_j\phi_k$.
Finally, the combination of $M^\ast$ and the lowest component of $\phi_i\phi_j\phi_k$
(that is, $A_iA_jA_k$) gives rise to the trilinear $A$-term.

As for anti-CASM ${\cal X}^+$, the dilaton is transformed into the antichiral compensator:
\be e^{2\varrho} \to (\chi^3)^+=
e^{2\varrho-2i\delta}[1+\sqrt{2}\bar\theta\Psi+\bar\theta\bar\theta M].\ee
Under this transformation, 
\bea \ln\mu^2 &\to & \ln\mu^2 -\ln(\chi^3)^+\nonumber\\
&=& \ln \mu^2-2\varrho+2i\delta -\sqrt{2}\bar\theta\Psi-\bar\theta\bar\theta M.\eea
Before closing the Appendix we would like to emphasize that this observation helps probe
the linear anomaly in Appendix D.

\section{The linear anomaly from the matter kinetic term}%

In this appendix we evaluate the LASM engendered by the kinetic term of the matter superfield
of the action (\ref{eq:sim.mssm}) (in momentum-space with Euclidean signature),
\be\label{eq:kin.mat} S \supset
\int \frac{d^\4 p}{(2\pi)^4}\,d^\2\theta d^\2\bar\theta\,{\phi_i}^{\!+}\phi_i.\ee
Utilizing dimensional regularization we evaluate the one-loop effective action from the diagram
in fig.~\ref{fig:D1}:
\be\label{eq:one.kin.mat} S_{\rm eff} = 
\int\frac{d^\4 p}{(2\pi)^\4}\,d^\2\theta d^\2\bar\theta\,{\phi_i}^{\!+}
(-p,\bar\theta)\phi_i(p,\theta)\,
\frac{1}{2}\frac{1}{16\pi^2}[y^\ast_{irs}y^{irs}-4g^2C_A(\phi_i)]\Sigma(p^2),\ee
where the factor of $1/2$ is a symmetric factor, $C_A(\phi_i)$ is the quadratic Casimir
constant of the representation of the gauge group to which the $\phi_i$'s belong:
\be (T^aT^a)_{jk}=C_A(\phi_i)\,\delta_{jk}\ee
(summing over $a$), and $\Sigma(p^2)$ is defined in Appendix A.
\FIGURE{\epsfig{file=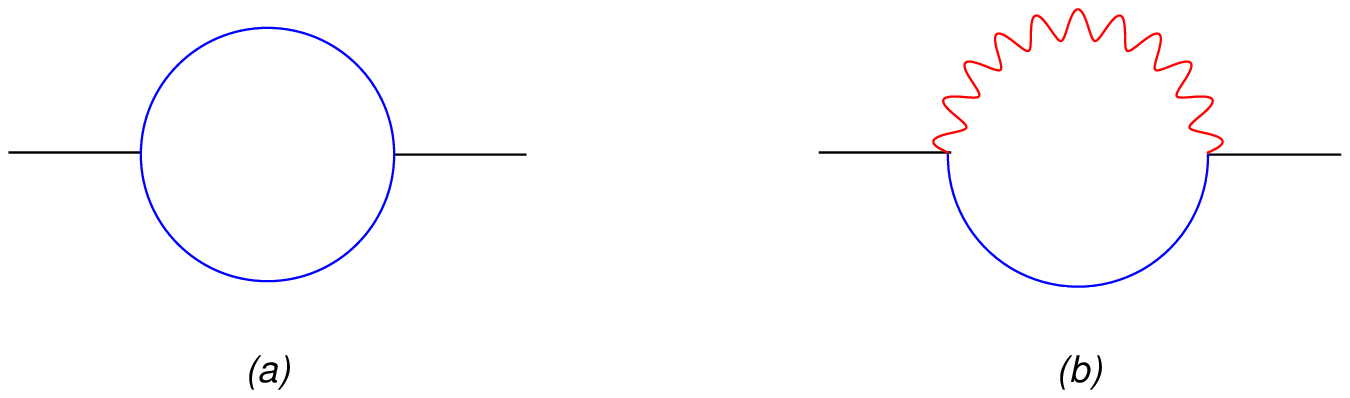,width=0.6\textwidth}
\caption{One-loop supergraphs contributing to the the ${\phi_i}^{\!+}\phi_i$
propagator.}\label{fig:D1}}

From (\ref{eq:kin.mat}) and (\ref{eq:one.kin.mat}), the wave function renormalization constant
$Z^{1/2}_i$ for the matter superfield $\phi_i$ is evaluated at one loop as
\be Z_i^j(\mu)=\delta^j_i-\frac{1}{16\pi^2}[y^\ast_{irs}y^{jrs}-4g^2\delta^j_iC_A(\phi_i)]
\frac{\mu^{2\varepsilon}}{2\varepsilon}, \ee
and the corresponding anomalous dimension (or gamma function) is immediately computed
(in the limit $\varepsilon\to 0$) as
\bea \gamma_i^j &\equiv& \frac{1}{2}\frac{\partial \ln Z_i^j}{\partial\ln\mu}=
\gamma_i^{(1)j}+\gamma_i^{(2)j}+\cdots,\\
\label{eq:gamma1}\gamma_i^{(1)j}&=&-\frac{1}{2(4\pi)^2}[y^\ast_{irs} y^{jrs}-4g^2\delta^j_i C_A(\phi_i)],\eea
where superscript $(n)$ denotes $n$-loop order.
Replacing $\ln\mu^2$ in (\ref{eq:one.kin.mat}) by 
$\ln\mu^2-\2\varrho-2i\delta-\sqrt{2}\theta\bar\Psi-\theta\theta M^\ast$,
we get nothing involved either in $\varrho$ or in $M^\ast$.
Here we assume that $\phi_i|_{\theta\theta}={\phi_i}^{\!+}|_{\bar\theta\bar\theta}=0$.
That is, supersymmetry is not spontaneously broken in the visible sector.
As a consequence, there is no one-loop contribution to the superconformal anomaly
associated with the matter kinetic term.

Next, we inspect the two-loop contribution to the superconformal anomaly.
First, the two-loop one-particle irreducible (1PI) supergraphs contributing to the 
${\phi_i}^{\!+}\phi_i$ propagator are shown in fig.~\ref{fig:D2},  
where a red blob denotes the relevant one-loop 1PI supergraph including any 
one-loop counter term that may be required~\cite{West:1984dg}.
Rather than directly calculate the two-loop 1PI supergraphs,
we simply quote the result from the ref.~\cite{West:1984dg,Jones:1983vk}.
The infinite part of the two-loop contribution to the two-point function $T_i^j$ is~\cite{Yamada:1994id}
\bea \label{eq:2looppro} T^j_i &=& \frac{-1+\varepsilon}{2(4\pi)^2\varepsilon^2}\,
\bigg\{4g^2_Ag^2_B C_A(\phi_i)C_B(\phi_i)\delta_i^j
+2g^4_A C_A(\phi_i)[\sum_i T_A(\phi_i)-3C_A(\mbox{adj.})]\delta_i^j\nonumber\\
&&\qquad\qquad +\,g^2_A[-C_A(\phi_i)+2C_A(\phi_l)]\,y^\ast_{ikl}\,y^{jkl}
-\frac{1}{2}\,y^\ast_{ikl}\,y^{lst}\,y^\ast_{qst}\,y^{jkq}\bigg\},
\eea
where we consider the general case with a semi-simple gauge groups 
$G=\Pi_A G_A$ with $G_A$'s being simple subgroups, and $g_A$ being 
the gauge coupling.
\FIGURE{\epsfig{file=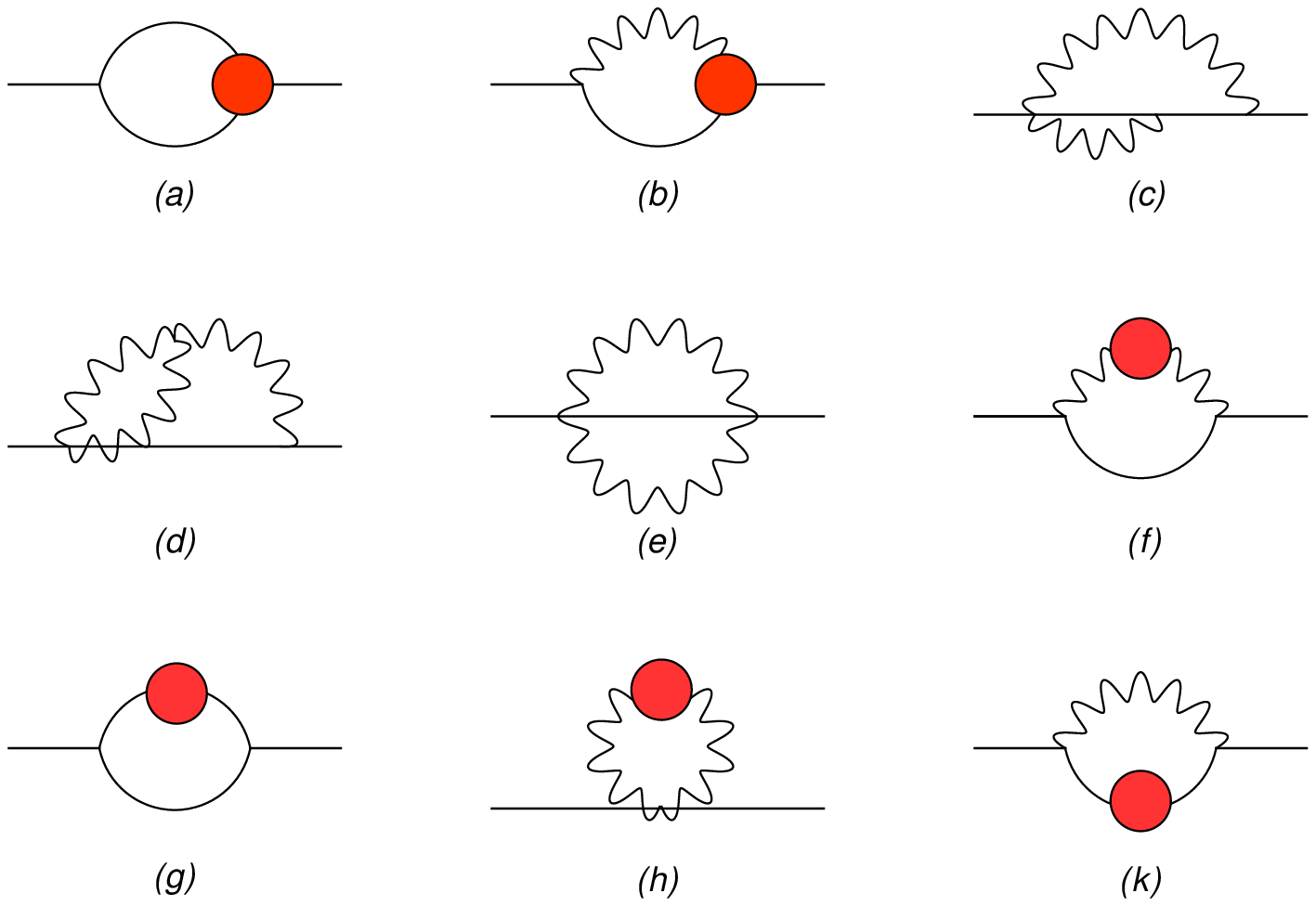,width=0.8\textwidth}
\caption{Two-loop 1PI supergraphs contributing to the ${\phi_i}^{\!+}\phi_i$
propagator. Each red blob denotes the relevant one-loop 1PI supergraph including any 
one-loop counter term.}\label{fig:D2}}
Comparing eq.~(\ref{eq:2looppro}) 
with the two-loop integral eq.~(3.14) in ref.~\cite{Abbott:1980jk},
we immediately find (see eq.~(2.11) in ref.~\cite{Martin:1993zk} or 
eq.~(25) in ref.~\cite{West:1984dg}.)
\bea\label{eq:two1pi}
\gamma_i^{(2)j} &=& -\frac{1}{(4\pi)^4}\,\bigg\{4g^2_Ag^2_B C_A(\phi_i)C_B(\phi_i)\delta_i^j
+2g^4_A C_A(\phi_i)[\sum_i T_A(\phi_i)-3C_A(\mbox{adj.})]\delta_i^j\nonumber\\
&&\,\,+\,g^2_A[-C_A(\phi_i)+2C_A(\phi_l)]\,y^\ast_{ikl}\,y^{jkl}
-\frac{1}{2}\,y^\ast_{ikl}\,y^{lst}\,y^\ast_{qst}\,y^{jkq}\bigg\},
\eea
which is the coefficient of the finite term proportional to $(\ln\mu^2)^2$.
What we need in two-loop contribution is indeed the finite term proportional to $(\ln\mu^2)^2$.
This is because $(\ln\mu^2)^2$ is replaced by
\be (\ln\mu^2)^2 \to 
(\ln\mu^2-\2\varrho-2i\delta-\sqrt{2}\theta\bar\Psi-\theta\theta M^\ast)
(\ln\mu^2-\2\varrho+2i\delta-\sqrt{2}\bar\theta\Psi-\bar\theta\bar\theta M).\ee
\FIGURE{\epsfig{file=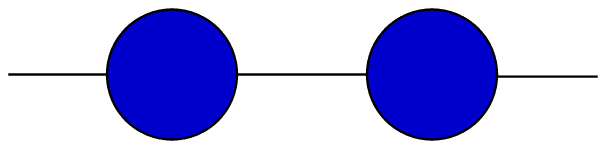,width=0.6\textwidth}
\caption{Two-loop one-particle reducible supergraph contributing to the ${\phi_i}^{\!+}\phi_i$
propagator. Each blue blob denotes either of the one-loop supergraphs 
shown in fig.~\ref{fig:D1}.}\label{fig:D3}}

Second, there are one-particle reducible supergraphs at two-loop, shown in fig.~\ref{fig:D3}.
Each blue blob shown in fig.~\ref{fig:D3} represents either of the one-loop supergraphs
shown in fig.~\ref{fig:D1}.
They should also be included in the computation for superconformal anomaly,
which is very clear from Appendix C where the one-loop one-particle reducible vertex supergraphs
make contributions to the superconformal anomaly but the one-loop 1PI vertex supergraphs do not.
From eq.~(\ref{eq:one.kin.mat}) and~(\ref{eq:gamma1}), we easily find 
that the coefficient of the finite term proprotional to $(\ln\mu^2)^2$ is
\be \label{eq:gamma11}
\gamma_i^{(1)l}\gamma_l^{(1)j}=\frac{1}{4(4\pi)^4}
[y^\ast_{irs} y^{lrs}-4g^2_A\delta_i^jC_A(\phi_i)]
[y^\ast_{lpq} y^{jpq}-4g^2_B\delta_i^jC_B(\phi_j)].
\ee
Adding eq.~(\ref{eq:two1pi}) and (\ref{eq:gamma11}) together we get the full expression 
proprotional to $(\ln\mu^2)^2$.
This fact is clear from the relation, up to two-loop order,
\be 
\gamma_i^{(2)j}+\gamma_i^{(1)l}\gamma_l^{(1)j}=-\frac{1}{2}\dot\gamma_i^j,\ee
with
\bea \dot\gamma_i^j&\equiv&\frac{\partial\gamma_i^j}{\partial\ln\mu}
=\frac{\partial\gamma_i^j}{\partial 
g}\beta_g+\frac{\partial\gamma_i^j}{\partial y^{krs}}\beta_{y^{krs}}, \\
\beta_g &\equiv& \frac{\partial g}{\partial \ln\mu},
\quad \beta_{y^{ijk}} \equiv \frac{\partial y^{ijk}}{\partial \ln\mu},\eea
which agrees with the previous results computed using spurion method~\cite{Pomarol:1999ie}.
Eventually the LASM is given as
\be L=-\frac{1}{2}\dot\gamma_i^i\,{\phi_i}^{\!+}\phi_i. \ee

\end{document}